# Shortest Paths in Less Than a Millisecond

Rachit Agarwal
University of Illinois
Urbana-Champaign, USA
agarwa16@illinois.edu

Matthew Caesar
University of Illinois
Urbana-Champaign, USA
caesar@illinois.edu

P. Brighten Godfrey
University of Illinois
Urbana-Champaign, USA
pbg@illinois.edu

Ben Y. Zhao
University of California
Santa Barbana, USA
ravenben@cs.ucsb.edu

## ABSTRACT

We consider the problem of answering point-to-point *shortest* path queries on massive social networks. The goal is to answer queries within tens of milliseconds while minimizing the memory requirements. We present a technique that achieves this goal for an extremely large fraction of path queries by exploiting the structure of the social networks.

Using evaluations on real-world datasets, we argue that our technique offers a unique trade-off between latency, memory and accuracy. For instance, for the LiveJournal social network (roughly 5 million nodes and 69 million edges), our technique can answer 99.9% of the queries in less than a millisecond. In comparison to storing all pair shortest paths, our technique requires at least 550× less memory; the average query time is roughly 365 microseconds — 430× faster than the state-of-the-art shortest path algorithm. Furthermore, the relative performance of our technique improves with the size (and density) of the network. For the Orkut social network (3 million nodes and 220 million edges), for instance, our technique is roughly 2588× faster than the state-of-the-art algorithm for computing shortest paths.

## Categories and Subject Descriptors

E.1 [**Data**]: Data Structures—*Graphs and networks*; H.2.4 [**Database Management**]: Systems—*Query processing*

## 1. INTRODUCTION

We consider the problem of answering point-to-point *shortest* path queries on massive social networks. Each query asks for the shortest path between a source-destination pair and the goal is to answer the query with minimal latency.

Point-to-point path queries are ubiquitous in social network industry. For instance, in professional networks like LinkedIn, it is desirable to find a short path from a job seeker to a potential employer; in social networks like Orkut, Facebook and Twitter, it is desirable to find how users are connected to each other; in social auction sites, distance and paths can be used to identify more trustworthy sellers [15]; academic networks (Microsoft academic search [6], for instance) compute paths between different authors; etc. More recently, these queries have also been used in the context of socially-sensitive and location-aware search [2,13], where it is required to compute distances (and paths) between a user and content of potential interest to the user. Many of these applications require (or can benefit from) computing *shortest* paths, which we focus on in this work.

Besides industry, point-to-point path queries are frequently used in research — to generate *unbiased samples* for distance-based graph analysis experiments [5, 11, 12, 16, 17, 19, 20], it is often desirable to obtain the shortest distance between each pair of nodes in a randomly sampled set of nodes.

Scaling point-to-point path queries to large scale social networks is challenging for two reasons. First, the latency requirements of the applications mentioned above are rather stringent — typically, it is desirable to answer queries within tens of milliseconds since higher latencies can be perceived by the users [10]. Such stringent latency requirements preclude the obvious option of running a shortest path algorithm for each query — a standard implementation of traditional shortest path algorithms even on relatively small networks (3 million nodes and 220 million edges) takes roughly 500 seconds on modern desktop computers [5].

Second, storing paths between each pair of users is infeasible due to memory limitations; even for a social network with 3 million users, this would require roughly 4.5 trillion entries. Social networks of interest, unfortunately, can be much larger in size — Facebook (800 million users), LinkedIn (135 million users), Twitter (200 million users), Orkut (more than 66 million users), etc.

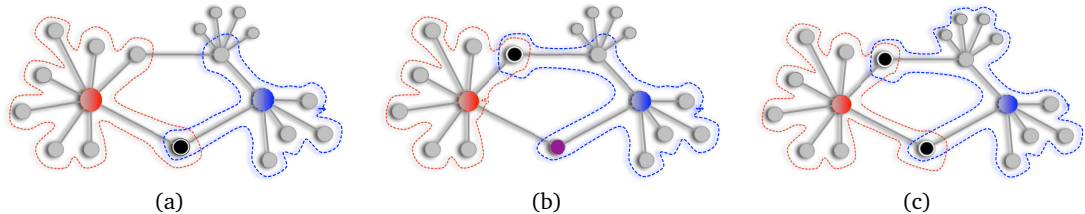

**Figure 1: High level idea of vicinity intersection.** The outlined area around the larger nodes denote their vicinities. (a) Extracting shortest paths and distances using vicinity intersection; (b) Selecting vicinities of fixed size may lead to incorrect results; (c) Selecting vicinities of fixed radius may require exploring a large fraction of the network.

There is a large body of work on point-to-point path query problem. We delay a complete discussion of related work to §4; however, we note that while heuristics like $A^*$ search [3,4] and bidirectional search [4] are useful in reducing the latency problem with traditional shortest path algorithms, they still require running a (modified) shortest path algorithm for each query and are unlikely to meet our latency requirements. For instance, our experiments (§3) show that bidirectional search can take hundreds of milliseconds to compute shortest paths even on moderate size networks. Citing lack of efficient techniques for computing *shortest* paths, a number of papers have developed techniques to compute *approximate* distances and paths [5, 11, 12, 16, 17, 19, 20] (see §4).

Our work differs from prior work in two main aspects. First, we focus on a much harder problem of computing *shortest* paths; and second, we exploit the structure of social networks to show that queries can be answered by exploring an extremely small fraction of the network. In particular, we use the idea of *vicinity intersection*, where the vicinity of a user is a (carefully defined) subset of users in its neighborhood. We observe that for any given pair of users: (1) if the vicinities intersect (have a user that lies in both the vicinities), one can extract the *shortest* path; and, (2) empirically, in social networks, the intersection nearly always[1] happens for vicinities of size roughly $c \cdot \sqrt{n}$ for some small $c$. Since each vicinity is an extremely small fraction of the entire network, each query can be answered quickly by exploring a vicinity. Our technique requires roughly $c \cdot n\sqrt{n}$ memory for storing vicinities; this may be significant for extremely large networks but does provide a reasonable trade-off between the two extreme solutions of storing all pair shortest paths (large memory) and online computation of shortest paths (large latency).

Using evaluations on real-world datasets, we argue that our technique offers a unique trade-off between latency, memory and accuracy. For instance, for the LiveJournal social network (5 million nodes and 69 million edges), our technique allows answering more than 99.9% of the queries by exploring less than 0.2% of the entire network. In comparison with storing all pair shortest paths, our technique requires at least a factor 550× less memory; each query can be answered in roughly 365 *microseconds* — a factor 430× faster than the state-of-the-art shortest path algorithm [4]. Furthermore, the relative performance of our technique improves with the size of the network. For the Orkut social network (3 million nodes and 220 million edges), for instance, our technique is a factor 2588× faster than the state-of-the-art algorithm [4].

[1] For source-destination pairs whose vicinities do not intersect, it is possible to combine our technique with those for computing exact [3,4] or approximate [5, 12, 17, 20] paths.

## 2. VICINITIES

We start the section by giving a high level description of our technique, intuitively defining vicinities and outlining some desirable properties of the vicinities (§2.1). We then formally define vicinities and give an algorithm to compute the vicinity of each node (§2.2). Finally, we confirm via empirical evaluation (using setup described in §2.3) that real-world social networks indeed exhibit the properties that we desire (§2.4).

### 2.1 High-level description of our technique

Our technique distributes the shortest path computations across two phases — an *offline* phase and an *online* phase. During the offline phase, for each node $u$ in the network, we compute and store information regarding a certain subset of nodes in the neighborhood of $u$, that we refer to as its vicinity. During the online phase, we use these vicinities to compute the shortest paths using the idea of vicinity intersection.

**What is vicinity intersection?** We say that the vicinities of a pair of nodes $s, t$ intersect if there is a node that lies in both the vicinities. We construct our vicinities such that if the vicinities of $s$ and $t$ intersect, one can retrieve the shortest path between $s$ and $t$. For instance, in Figure 1(a), the vicinities of the larger nodes (denoted by the outlined area) intersect at the black (lower middle) node, and indeed, this node lies along the shortest path; by combining the respective paths from larger nodes to the black node, the query can be answered. Finding such a node requires iterating through each node in the vicinity of the source and checking whether it is contained in the vicinity of the destination (while keeping track of the shortest path found so far).

**How do we *not* define vicinities?** Consider the following two strawman definitions: (1) a fixed number of closest nodes; and (2) all the nodes within some fixed distance. Figure 1(b) shows an example where the first definition leads to incorrect results. In this example, we let the vicinity of each node to be its 8 closest neighbors. Then, if ties are broken arbitrarily, vicinities intersect (at black node along a path of length 3 hops) but not along the shortest path (which is of 2 hops).

Figure 1(c) shows an example where the second definition leads to inefficiency. Indeed, some nodes may have a very large number of nodes within a given distance due to having *dense neighborhoods*; then, by this strawman definition, each of these nodes will have a large number of nodes in its vicinity. Since checking vicinity intersection requires iterating through each node in the vicinity, this may lead to high latency. Furthermore, many nodes having dense neighborhoods also leads to high memory requirements.

**What properties do we want from the vicinities?** Based on the discussion above, there are three desirable properties of vicinities. First, vicinities must guarantee correctness; that is, if vicinities of $s$ and $t$ intersect, one of the nodes that lies in this intersection must be on the shortest path between $s$ and $t$. Second, vicinities should be large enough so that most of the source-destination pairs have intersecting vicinities; it is also desirable that vicinities can be computed rather efficiently. Finally, vicinities should be small enough so that the memory requirements are reasonable (recall that our technique requires storing, for each node $u$ in the network, information regarding each node in the vicinity of $u$).

**How do we define vicinities?** We give an intuitive description of the notion of vicinities for the case of unweighted networks. We first construct a set of nodes (denoted by $L$) by sampling each node in the network with a probability proportional to its degree. Informally speaking, we define the vicinity of each node $u$ (for unweighted networks) to be the set of nodes $v$, such that distance between $u$ and $v$ is no more than the distance from $u$ to its closest node in $L$.

Such a construction gives vicinities that possess all the properties discussed above. First, the vicinity of each node $u$ contains all nodes that are within certain distance from $u$ (that is, there is no tie-breaking); we show that this is sufficient to guarantee the correctness property. Second, by setting the sampling probability for construction of set $L$ appropriately, we ensure that vicinities are large enough so that most source-destination pairs have intersecting vicinities. Finally, our sampling algorithm for construction of set $L$ ensures that nodes that have extremely dense neighborhoods avoid having extremely large vicinities. Intuitively, a node $u$ that has dense neighborhood is likely to have a high degree node (say, $\ell(u)$) in its neighborhood; by way of constructing $L$, we ensure that $\ell(u)$ has high likelihood to be in $L$ and hence, the vicinity of $u$ stops "expanding" after hitting $\ell(u)$. This allows us to ensure that vicinities are small enough to bound the memory requirements.

## 2.2 Definition and construction algorithm

We now formally define vicinities and sketch an algorithm that efficiently constructs the vicinity of each node. We use the notation described in Table 1. We assume that each edge in the network is assigned a non-negative weight; for unweighted networks, this weight is assumed to be 1.

**Formal definition of vicinities and boundary nodes.** Let $L$ be a subset of nodes constructed by sampling each node $u$ with a probability proportional to the degree of $u$ (we describe the exact expression for the probability in the construction algorithm below). Then, given this set $L$, the vicinity for each node is defined as follows (see notation in Table 1):

DEFINITION 1. *For any undirected network $G = (V,E)$, the ball of a node $u \in V$, denoted by $B(u)$, is the set of nodes $v \in V$ for which $d(u,v) < d(u,\ell(u))$. The vicinity of $u$, denoted by $\Gamma(u)$, is the set of nodes in $B(u) \cup N(B(u))$.*

We will later use the idea of boundary nodes to optimize our implementation of vicinity intersection. For any node $u$ and a node $v \in \Gamma(u)$, we say that $v$ belongs to the **boundary** of $u$ if $N(v) \nsubseteq \Gamma(u)$, that is, has links to nodes outside the vicinity of $u$. For any node $u$, we denote the set of nodes on the boundary by $\mathcal{B}(\Gamma(u))$. The **boundary size** of any node $u$ is then denoted by $|\mathcal{B}(\Gamma(u))|$.

**Table 1: Notation used in the paper.** $G = (V,E)$ **is assumed to be a connected, undirected network;** $V'$ **and** $L$ **are subset of nodes in** $V$.

| | |
|---|---|
| $n$ | number of nodes in $G$ |
| $m$ | number of edges in $G$ |
| $N(u)$ | set of all the neighbors of $u$ |
| $N(V')$ | set of all neighbors of nodes in $V'$ |
| $d(u,v)$ | distance between $u$ and $v$ in $G$ |
| $\ell(u)$ | the node $a \in L$ that minimizes $d(u,a)$ |
| | (ties broken arbitrarily) |
| $\Gamma(u)$ | vicinity of node $u$ (Definition 1) |
| $\mathcal{B}(\Gamma(u))$ | boundary nodes of $u$ |
| $\alpha$ | a parameter controlling vicinity size |

**An algorithm for constructing the vicinities.** The construction algorithm takes as input a parameter $\alpha$ that controls the size of the vicinities. In the first step, each node $u$ in the network is sampled with a probability

$$p_s = \frac{m}{\alpha n \sqrt{n}} \cdot \left\lceil \frac{2n}{m} \cdot \deg(u) \right\rceil$$

Denote the set of sampled nodes as $L$. The algorithm then constructs the vicinity of each node (as defined in Definition 1) — starting at each node $u$, it runs a modified shortest path algorithm [16] that stops once all the nodes at distance $d(u,\ell(u))$ or less have been visited. All these nodes constitute the vicinity of $u$, as in Definition 1. It is easy to prove that if the set $L$ is constructed by sampling each node with a probability $p_s$ as above, the expected size of the vicinity of each node is $\alpha \cdot \sqrt{n}$.

## 2.3 Datasets and Experimental Setup

We outlined a set of desirable properties of the vicinities in §2.1. In this subsection, we describe our experimental setup used to study these properties empirically using social network datasets. The datasets used in our experiments are shown in Table 2. The DBLP dataset is from [18]; the LiveJournal dataset is from [14] and the rest of the datasets are from [9].

**Table 2: Social network datasets used in evaluation.**

| Topologies | # Nodes (Million) | # Directed Links (Million) | # Undirected Links (Million) |
|---|---|---|---|
| DBLP | 0.71 | 2.51 | 2.51 |
| Flickr | 1.72 | 22.61 | 15.56 |
| Orkut | 3.07 | 223.53 | 117.19 |
| LiveJournal | 4.85 | 68.99 | 42.85 |

For each dataset, we did the following set of experiments — we constructed vicinities of size $\alpha \cdot \sqrt{n}$ using the algorithm in §2.2 for $\alpha$ varying from 1/16 to 16; sampled 1000 random nodes; and, checked for every pair of sampled nodes (resulting in 1 million source-destination pairs per experiment), whether or not their vicinities intersect. For each dataset, we repeated the experiment 10 times, resulting in roughly 10 million unbiased samples.

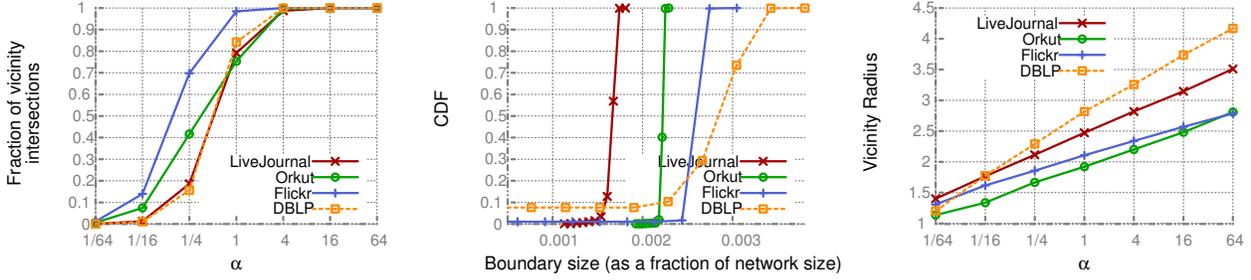

**Figure 2:** Properties of the vicinities; recall, $\alpha$ is the parameter that controls the size of the vicinities. (left) Most vicinities intersect for $\alpha \geq 4$; (center) Vicinities have small number of boundary nodes for $\alpha = 4$; (right) Average vicinity radius is small even for $\alpha = 4$.

## 2.4 Properties of vicinities

We confirm, via empirical evaluation on real-world datasets, that the vicinities admit the desirable properties outlined in §2.1; we also present two additional properties that are later used as optimizations.

We start with the correctness property; recall, we want to prove that if the vicinities of any pair of nodes $s, t$ intersect, then one of the nodes in the intersection must lie along the shortest path. We prove this property in Appendix A (in fact, we prove a stronger property which shows that it suffices to check for intersection of boundary of $s$ and $t$). We now return to the other properties.

**For $\alpha = 4$, vicinities have (small) bounded size and can be constructed efficiently.** We give a sketch of the proof. By sampling each node with a probability $p_s$ (from §2.2), one can prove that the size of set $L$ is roughly $\frac{m}{\alpha\sqrt{n}}$. It is rather easy to see (using [16, Lemma 3.2] and [1, Section 5]) that the vicinity of each node is of size roughly $\alpha \cdot \sqrt{n}$. Furthermore, each vicinity can be computed in time $O(\alpha \cdot \sqrt{n})$ using the modified shortest path algorithm presented in [16].

**For $\alpha = 4$, any two vicinities intersect with high probability.** Figure 2(a) shows the variation of fraction of vicinity intersections (averaged over all source-destination pairs over all experiments) with $\alpha$. The figure shows that for $\alpha = 4$, the vicinities of any two randomly selected nodes intersect with an extremely high probability. In fact, for all datasets, a value of $\alpha = 16$ suffices to achieve vicinity intersection for each source-destination pair.

**For $\alpha = 4$, vicinities have small boundary size.** Figure 2(b) shows the CDF (over sampled nodes) of the boundary size as a fraction of the number of nodes in the network for the case of $\alpha = 4$. We note that, in the worst-case, vicinities have boundary size of less than 0.4% of the nodes in the network.

**For $\alpha = 4$, vicinities have small radius.** Let the vicinity radius of any node $u$ be defined as $d(u, \ell(u))$. Figure 2(c) shows the variation of vicinity radius (averaged over all nodes) with $\alpha$. Note that for $\alpha = 4$, the vicinity radius is less than 3.5 hops on an average.

We will use the second property to bound memory and latency requirements; the third property to argue about the accuracy and the remaining properties to argue about the performance of our technique in terms of latency.

## 3. COMPUTING SHORTEST PATHS

In this section, we formally describe our algorithm (§3.1). We also show, using preliminary evaluation results (§3.2), that even with a straightforward implementation, our technique can compute shortest paths in less than a millisecond even for networks over 5 million nodes and 69 million edges.

### 3.1 The Technique

In this subsection, we formally describe our algorithm and prove its correctness. We start by describing the algorithm for a simpler problem — retrieving the exact distance between the queried nodes. We then extend the algorithm to retrieve the corresponding path.

Our data structure stores, for each node $u$, a hash table containing the exact distance to each node $v \in \Gamma(u)$. In addition, if $u \in L$, the data structure stores a hash table containing the exact distance from $u$ to each other node $v \in V$. When queried for distance between $s$ and $t$, the exact distance is returned (directly using one of the stored hash tables) if either of the following four conditions is satisfied: (1) $s \in L$; (2) $t \in L$; (3) $s \in \Gamma(t)$; or, (4) $t \in \Gamma(s)$; if not, the algorithm performs vicinity intersection — it checks for each node $v \in \Gamma(s)$ whether $v \in \Gamma(t)$. The algorithm also keeps track of the minimum path length seen among all $v \in \Gamma(s) \cap \Gamma(t)$; once the vicinity intersection check is complete, this minimum value is guaranteed to be the shortest path length.

The algorithm, as described above, iterates through each node in $\Gamma(s)$ (or $\Gamma(t)$) and hence, requires as many checks as the number of nodes in the vicinity of $s$ (or $t$). Our second observation is that vicinity intersection check for a given pair of nodes $s$ and $t$ can be performed by iterating through the boundary nodes of either $s$ or $t$. Since $\mathcal{B}(\Gamma(s)) \subseteq \Gamma(s)$, this may require fewer checks and may result in reduced latency. The final algorithm is shown in Algorithm 1. We prove its correctness in the Appendix.

---

**Algorithm 1** Vicinity-intersection(u, v)

1: **Input:** NODES $s, t$, VICINITIES $\Gamma(s), \Gamma(t)$
2: $\quad \delta \leftarrow \infty$
3: $\quad$ If $s \in L$ OR $t \in L$ OR $t \in \Gamma(s)$ OR $s \in \Gamma(t)$
4: $\quad\quad$ Return $d(s, t)$
5: $\quad$ For each $w \in \mathcal{B}(\Gamma(s))$
6: $\quad\quad$ If $w \in \Gamma(t)$
7: $\quad\quad\quad$ If $d(s, w) + d(t, w) < \delta$
8: $\quad\quad\quad\quad \delta = d(s, w) + d(t, w)$
9: $\quad$ Return $\delta$

Table 3: Query time results for various datasets for $\alpha = 4$.

| Dataset | Our technique | | | BFS | Bidirectional BFS | Speed-up |
| --- | --- | --- | --- | --- | --- | --- |
| | # Hash-table look-ups | | Time (in ms) | Time (in ms) | Time (in ms) | (compared to |
| | average-case | worst-case | | | | Bidirectional BFS) |
| DBLP | 1847.12 | 2124 | 0.094 | 327.2 | 18.614 | 198× |
| Flickr | 4898.78 | 5067 | 0.228 | 2090.2 | 83.956 | 368× |
| Orkut | 6877.52 | 6937 | 0.294 | 28678.5 | 760.987 | 2588× |
| LiveJournal | 8185.71 | 8360 | 0.363 | 6887.2 | 156.443 | 431× |

We now extend the algorithm to generate the paths. In order to do so, we first modify our data structure slightly — for each node $u$ and each node $v \in \Gamma(u)$, our data structure will now store the next hop to $v$ (recall, for distance computations, we simply stored the distance to $v$). The algorithm then returns the paths as follows: first, the node $v_{i_0}$ along the shortest path is identified; the path is then retrieved by following the series of next-hops until we reach $v_{i_0}$.

### 3.2 Performance

The results from §2.4 imply that by using roughly $4\sqrt{n}$ memory per node, our technique computes shortest paths for more than 99.9% of the source-destination pairs. Our technique, hence, requires $\sqrt{n}/4$ factor less memory when compared to storing all-pair shortest paths, at the expense of less than 0.1% loss in accuracy. We now present the latency results using a preliminary evaluation.

Our current implementation stores the vicinities of nodes in-memory using hash tables (unordered map); the hash table implementation is as provided by the GNU C++ STL. The implementation runs on a single core of a Core i7-980X, 3.33 GHz processor running Ubuntu 10.10 with Linux kernel 2.6.35-32. Table 3 compares the query time of our technique with that of an optimized implementation of breadth-first algorithm and bidirectional breadth-first algorithm [4] using the set-up described in §2.4 (for $\alpha = 4$; this is the case when more than 99.9% of the vicinities pair intersect).

We make two observations. First, our technique is at least two orders of magnitude faster than the best known algorithm even for small networks. Second, the relative performance of our technique improves with the size of the network. Intuitively, the number of operations performed to compute the shortest path increase with the number of nodes (both for our technique and for [4]); each such operation has higher latency for shortest path algorithms when compared to our technique (we need hash table look-ups while [4] requires priority queue operations, etc.). Furthermore, the latency of our technique has low dependency on the density of the network; on the other hand, latency of shortest path algorithm increases significantly with little increase in density of the network.

## 4. RELATED WORK

Our goals are related to two key areas of related work:

**Shortest path algorithms and heuristics.** Heuristics like $A^*$ search [3,4] and bidirectional search [4] have been proposed to overcome the latency problems with traditional algorithms for computing shortest paths. The approaches in [3, 4], although useful in reducing the query time, still require running a (modified) shortest path algorithm for each query and do not meet the latency requirements. For instance, the experimental results in §3 shows that bidirectional search can take hundreds of milliseconds to compute the shortest paths even on moderate size networks.

In comparison to [3, 4], our contributions are two-fold: first, we show that empirically, in social networks, vicinities of size $4\sqrt{n}$ nearly always intersect (heuristics in [3,4] could also exploit this); and second, we argue that the vicinities being a small fraction of the entire network, storing and checking intersection quickly is feasible. This should be substantially faster than traditional bidirectional search [4] because it is just a series of hash table look-ups in a relatively compact data structure with one element per vicinity node — as opposed to running a shortest path algorithm that would require priority queue operations, and may even explore a large fraction of the entire network.

**Approximation algorithms.** Arguing that the above heuristics [3, 4] are unlikely to meet the stringent latency requirements of social network applications, [5, 11, 12, 17, 19, 20] focus on computing *approximate* distances and paths. Our work differs in two ways. First, unlike [11,19], our technique returns the actual paths. Second, we can *efficiently* compute shortest paths for a large fraction of source-destination pairs; prior techniques that have comparable latency [12] return paths that have an absolute error of more than 3 hops on an average and techniques that have comparable accuracy [5,17,20] have a latency of tens to hundreds of milliseconds (see [20] for a detailed comparison).

## 5. RESEARCH CHALLENGES

In terms of applicability of our technique, we see two challenges that need to be resolved. First, our current evaluations are on networks of up to 5 million nodes (LiveJournal) and up to 220 million edges (Orkut); do the results hold for larger networks? The results from §3.2 suggest that the relative performance of our technique should improve for larger networks; confirming this is the focus of our ongoing work. Second, is it possible to extend our approach to social networks modeled as directed networks (Twitter, for example)?

In terms of performance, several challenges remain. First, our technique requires storing roughly $4\sqrt{n}$ memory per node; while reasonable, is it possible to reduce the memory requirements while maintaining the current accuracy and latency? Second, can we further reduce the latency of our technique using more customized implementations of the data structures? Finally, shortest path queries are notoriously hard to parallelize [7] requiring either large memory at each machine (to replicate the input network across each machine) or large amounts of data transfer [8]. Is it possible to parallelize our technique without replicating the data structure?

## Acknowledgments
This work was in part supported by National Science Foundation grants CNS-1053781 and CCF-1017069 and in part by DARPA GRAPHS (BAA-12-01).

## APPENDIX
## A. PROOF OF CORRECTNESS

We start by proving that if the vicinities of a pair of nodes $s, t$ intersect, one of the nodes that lies in the intersection must belong to the shortest path between $s$ and $t$. We prove the correctness for unweighted graphs; the proof for weighted graphs is a simple extension of this proof.

THEOREM 1. *For any pair of nodes $s, t$, if $\Gamma(s) \cap \Gamma(t) \neq \emptyset$, then there exists a node $w \in \Gamma(s) \cap \Gamma(t)$ such that $w$ lies on the shortest path between $s$ and $t$.*

PROOF. Let $w \in \Gamma(s) \cap \Gamma(t)$ be the node that minimizes $d(s, w) + d(t, w)$. For sake of contradiction, assume that $w$ does not lie along the shortest path between $s$ and $t$; that is, $d(s, t) < d(s, w) + d(w, t)$.

Let $P = (s, v_0, \ldots, v_k, t)$ be the shortest path between $s$ and $t$ and let $i_0 = \max\{i : v_i \in P \cap \Gamma(s)\}$. Note that $v_{i_0} \notin \Gamma(t)$, else by setting $w = v_{i_0}$, we will contradict the assumption that $w$ does not lie along the shortest path between $s$ and $t$. By definition of the vicinity, for an unweighted graph, we have that $d(s, v_{i_0}) \geq d(s, w)$ since otherwise $v_{i_0+1}$ must be in $\Gamma(s)$. Furthermore, since $v_{i_0} \notin \Gamma(t)$, we have that $d(t, v_{i_0}) > d(t, w)$. Hence, we get that $d(s, t) = d(s, v_{i_0}) + d(t, v_{i_0}) > d(s, w) + d(t, w)$, leading to the contradiction that $P$ is the shortest path. □

Next, we show the correctness of the optimization used in Algorithm 1 — if the vicinities of two nodes intersect, the intersection occurs at one of the boundary nodes; hence, it suffices to iterate through the boundary nodes to check for vicinity intersection.

LEMMA 1. *For any pair of nodes $s, t$, consider the case when $s \notin \Gamma(t)$ and $t \notin \Gamma(s)$. Then, we have that $\Gamma(s) \cap \Gamma(t) = \emptyset$ if and only if $\mathcal{B}(\Gamma(s)) \cap \Gamma(t) = \emptyset$.*

PROOF. ($\Rightarrow$) This is trivially proved using the observation that for any node $s$, we have that $\mathcal{B}(\Gamma(s)) \subseteq \Gamma(s)$.

($\Leftarrow$) Suppose $\Gamma(s) \cap \Gamma(t) \neq \emptyset$; we show that this implies $\mathcal{B}(\Gamma(s)) \cap \Gamma(t) \neq \emptyset$. Assume that the condition in the statement of the lemma holds; that is, $s \notin \Gamma(t)$ and $t \notin \Gamma(s)$. Then, by Theorem 1, there exists a node $w \in \Gamma(s) \cap \Gamma(t)$ that lies along the shortest path between $s$ and $t$. Denote by $P(s, t) = (s, v_0, \ldots, v_k, t)$ the shortest path between $s$ and $t$ and let $i_0 = \max\{i : v_i \in P \cap \Gamma(s)\}$. Then, clearly, $v_{i_0} \in \Gamma(s) \cap \Gamma(t)$. Furthermore, since $v_{i_0+1} \notin \Gamma(s)$, we have that $v_{i_0} \in \mathcal{B}(\Gamma(s))$. The proof follows by noting that $v_{i_0} \in \Gamma(t)$. □